\documentclass[journal,twoside,web]{ieeecolor}
\usepackage{generic}
\usepackage{amsmath,amssymb,amsfonts}
\usepackage{algorithmic}
\usepackage{cite}
\usepackage{graphicx}
\usepackage{algorithm,algorithmic}
\usepackage{hyperref}
\hypersetup{hidelinks=true}
\usepackage{textcomp}
\def\BibTeX{{\rm B\kern-.05em{\sc i\kern-.025em b}\kern-.08em
    T\kern-.1667em\lower.7ex\hbox{E}\kern-.125emX}}
\RequirePackage{etex} 

\usepackage{generic}

\usepackage{amsmath}

\allowdisplaybreaks
\usepackage{xcolor}
\usepackage{footnote}
\usepackage{algorithm}
\usepackage{algorithmic}
\usepackage{multirow} 
\usepackage{booktabs}
\usepackage{graphicx}
\usepackage{amssymb}
\usepackage{amsbsy}

\usepackage{amsthm}

\usepackage{array}
\usepackage{longtable}
\usepackage{epstopdf}
\usepackage{pbox}
\usepackage{breqn}
\usepackage{mathrsfs}
\usepackage{multicol}
\usepackage{supertabular}
\usepackage{enumerate}
\usepackage[colorinlistoftodos]{todonotes}
\usepackage{url}
\usepackage[justification=centering]{caption}
\usepackage{tabu}
\usepackage{subcaption}
\usepackage{graphics} 
\usepackage{epsfig} 

\usepackage{mathtools}

\usepackage{tikz}
\usepackage{tikz-network}

\usepackage{cases}
\usepackage{bm}

\usepackage{aligned-overset}

\newtheorem{thm}{Theorem}
\newtheorem*{thm*}{Theorem}

\newtheorem{lmm}{Lemma}
\newtheorem{asm}{Assumption}

\theoremstyle{definition}

\newtheorem{dfn}{Definition}

\newcommand{\thistheoremname}{}
\newtheorem*{genericthm*}{\thistheoremname}
\newenvironment{namedthm*}[1]
  {
  \renewcommand{\thistheoremname}{#1}%
  \begin{genericthm*}}
  {\end{genericthm*}}

\usepackage{xcolor}

\usepackage{amsmath}
\begin{document}
\title{
Learning-Based Cost-Aware Defense of Parallel Server Systems against Malicious Attacks
}

\author{Yuzhen Zhan and Li Jin
\thanks{This work was in part supported by NSFC Project 62473250, SJTU UM Joint Institute, and J. Wu \& J. Sun Endowment Fund.}
\thanks{The authors are with the UM Joint Institute, Shanghai Jiao Tong University, China. (Emails: zyzhen1@sjtu.edu.cn, li.jin@sjtu.edu.cn.)}
}
\maketitle

\begin{abstract}
We consider the cyber-physical security of parallel server systems, which is relevant for a variety of engineering applications such as networking, manufacturing, and transportation. These systems rely on feedback control and may thus be vulnerable to malicious attacks such as denial-of-service, data falsification, and instruction manipulations. In this paper, we develop a learning algorithm that computes a defensive strategy to balance technological cost for defensive actions and performance degradation due to cyber attacks as mentioned above. 
We consider a zero-sum Markov security game. We develop an approximate minimax-Q learning algorithm that efficiently computes the equilibrium of the game, and thus a cost-aware defensive strategy. 
The algorithm uses interpretable linear function approximation tailored to the system structure.
We show that, under mild assumptions, the algorithm converges with probability one to an approximate Markov perfect equilibrium.
We first use a Lyapunov method to address the unbounded temporal-difference error due to the unbounded state space. We then use an ordinary differential equation-based argument to establish convergence.
Simulation results demonstrate that our algorithm converges about 50 times faster than a representative neural network-based method, with an insignificant optimality gap between 4\%--8\%, depending on the complexity of the linear approximator and the number of parallel servers.

\end{abstract}


\section{Introduction}

\subsection{Motivation}
Modern parallel server systems, such as cloud computing \cite{laszka2019detection}, industrial production lines \cite{fraile2018trustworthy}, and intelligent transportation networks \cite{jin2018stability}, rely on dynamic routing to optimize performance (delay and throughput). 
However, routing performance depends on connected and autonomous components subject to inherent cyber-physical security risks, especially in the face of increasingly sophisticated malicious cyberattacks \cite{hu2018resilient}.
Cyberattacks including Denial of Service (DoS), data falsification, and routing deception can severely compromise system performance \cite{liu2021resilient, lucia2022supervisor,cao2022sliding}. 
In transportation, falsifed traffic information can mislead vehicles and cause congestion \cite{feng2022cybersecurity}. 
For web services, a falsified blocking report in a web server farm could redirect traffic to an already overloaded server \cite{aslam2024scrutinizing}. Similar risks plague production lines and communication networks, where strategic attacks can inject erroneous instructions to disrupt critical operations \cite{achleitner2016cyber}. Actual incidents have been reported \cite{news1,news2,news3}.
Given the criticality of these systems in modern society, it is essential to address the potential physical performance degradation due to cyber security risks \cite{van2019internet}. 
\captionsetup[figure]{font=small}
   \begin{figure}[thpb]
      \centering
      \includegraphics[width=0.48\textwidth]{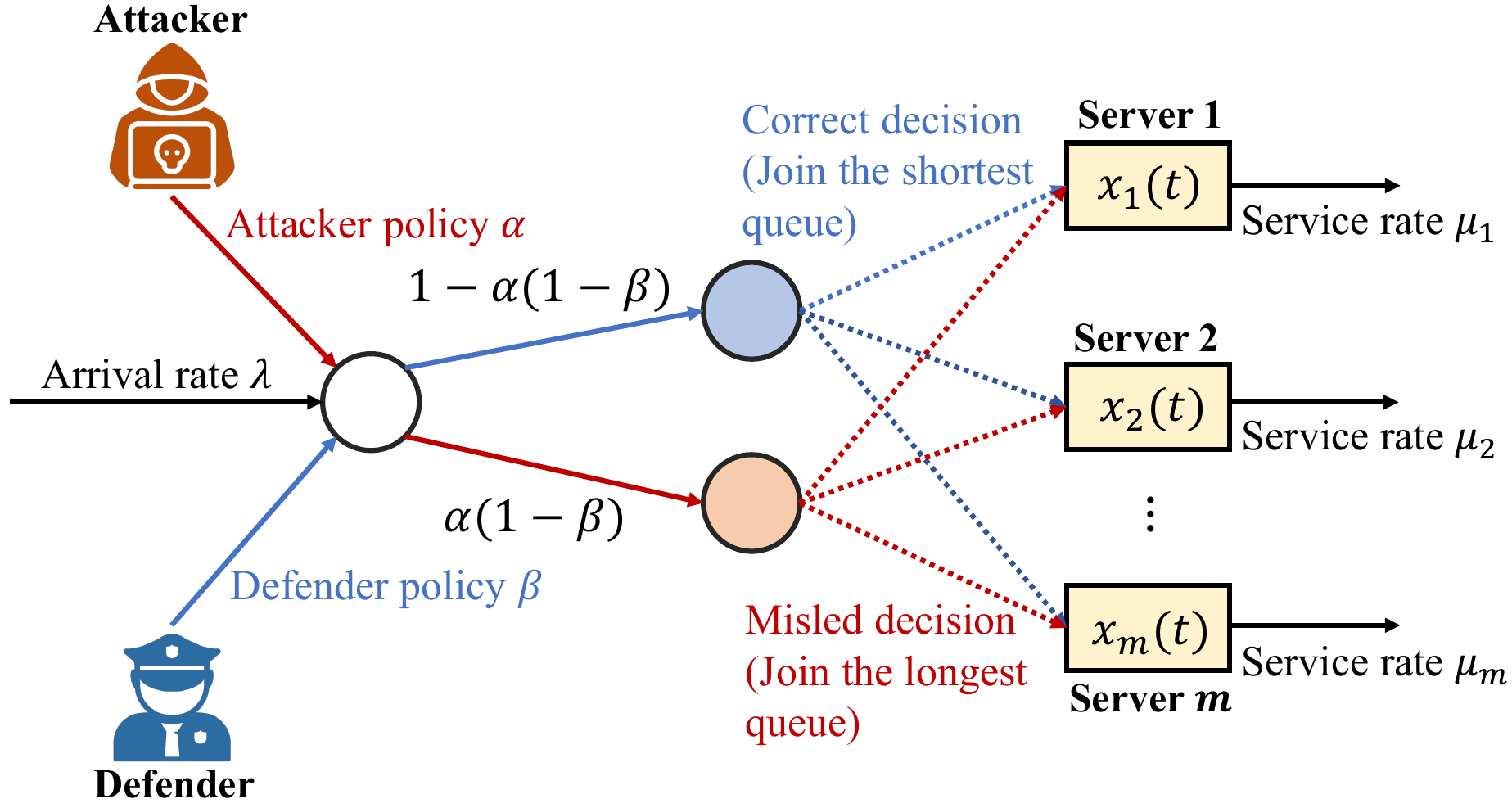}
      \caption{An $m$-server system with join-the-shortest-queue routing subject to security failures.}
      \label{fig_system}
   \end{figure}
   
Game theory has emerged as a powerful analytical framework for understanding cyber-physical security threats, particularly for strategic interactions between attackers and defenders \cite{tushar2023survey}.
By modeling the adversarial relationship as a game, a defender can anticipate attack strategies and optimize resource allocation to reduce potential losses \cite{xie2024cost}. Static security game models have been successfully used to study long-term trade-offs between security investments and defense effectiveness \cite{liang2012game,khouzani2011optimal}. 
Dynamic/stochastic game models such as the one considered in this paper are more suitable for real-time attacker-defender interactions \cite{wu2019securing,xie2024cost,miao2018hybrid}. However, these models usually have complex dynamics. Consequently, computational complexity is a critical bottleneck, which limits their applicability in realistic engineering systems. 

One promising solution to the above challenges is reinforcement learning (RL) methods \cite{lian2021inverse, xu2024cooperative}. 
RL methods are capable of computing optimal/equilibrium strategies in complex Markov games with unknown model information \cite{xie2020learning,zhou2021robust,bozkurt2024learning}.
Since we consider parallel server systems with unbounded state spaces, we will use value function approximation.
However, the broad class of neural network-based methods usually lack theoretical guarantees on training convergence and system stability \cite{bucsoniu2011approximate,ma2023safe}, which are essential for cyber-physical security analysis.
Existing convergence guarantees for approximate methods are mostly developed for finite-state or bounded-state problems; very limited results have been developed for games with unbounded state spaces \cite{prieto2015approximation}. 
The key challenge for unbounded state spaces is that the $\infty$-norm-based argument used for finite/bounded problems does not directly apply.


\subsection{Related work}


For Markov decision processes with unbounded state space but with bounded reward, learning methods have been understood fairly well \cite{shah2020stable,ding2022independent}. 
For problems with unbounded reward, most studies considered convergence guarantees for RL methods under fairly stringent conditions, such as rapidly decaying discount factors \cite{chen2022finite}, implicitly modified discount rates \cite{zhang2021breaking}, and linear transition dynamics or linear reward \cite{zhou2021provably}. 
In particular, Melo et al. \cite{melo2007convergence} proposed a set of milder conditions under which Q-learning with linear function approximation converges with probability one, which provides a solid foundation for our analysis.

For games with finite/bounded state spaces, extensive learning methods have been developed and studied
\cite{sayin2021decentralized,ozdaglar2021independent}. In particular, Littman \cite{littman1994markov} proposed the minimax Q (MQ) learning algorithm for finite-state games; our algorithm builds on this classical baseline. Szepesv\'ari and Littman \cite{szepesvari1999unified} 
established convergence guarantees for this algorithm.
Hu et al. \cite{hu2003nash} introduced Nash Q-learning, extending MQ to general-sum games. 
Lowe et al. \cite{lowe2017multi} extended MQ to cooperative-competitive environments by incorporating neural networks.  
Recently, Chen et al. \cite{chen2023two} extended the MQ learning to payoff-based scenarios with function approximation; although this work considers bounded state spaces, it gives useful insights for unbounded settings. 


For games with unbounded state spaces, very limited learning-based methods have been developed. For such games, existing results typically focus on dynamic programming methods and the existence of a stationary equilibrium \cite{prieto2015approximation,guo2005nonzero}. A typical method to address unboundedness is to use Lyapunov functions, which is commonly used in for control of queuing systems \cite{kumar1995stability,xie2022stabilizing, xie2024cost}. 
However, Lyapunov methods are more suitable for traffic stabilization but are less helpful for equilibria computation.

A standard approach to convergence analysis of stochastic approximation is the the ordinary differential equation (ODE) method. The approach was formalized and extended by Borkar and Meyn \cite{borkar1998stability}. Recent advancements such as quasi-stochastic approximations and dynamical systems views have improved its convergence rate and adaptation to more general noise conditions \cite{chen2021revisiting,benveniste2012adaptive}. However, current ODE-based frameworks often assume global Lipschitz continuity and noises with bounded variance—conditions that are relatively strict in practical scenarios \cite{liu2025ode}. 
The key to establish convergence of MQ learning in our setting is to jointly study the behavior of the traffic state and the approximate Q function.
To the best of our knowledge, this problem has not been fully understood, especially in a game-theoretic setting. 

\subsection{Our contributions}

In this paper, we develop an approximate minimax-Q (AMQ) learning algorithm to compute a near-equilibrium cost-aware defensive strategy for parallel server systems subject to malicious attacks (Fig.~\ref{fig_system}). This algorithm extends the classical MQ learning algorithm to the unbounded setting of parallel server systems. We use linear value function approximation, with a rather broad class of bases, to cover the unbounded state space. Importantly, we design the structure of the approximate Q function with insights about the systems dynamics, which makes the weights interpretable. We also provide a Foster-Lypaunov drift-based qualification for the behavior policy. We show that a qualified behavior policy must exist if the system is stabilizable.


The main result (Theorem 1) states that the proposed AMQ algorithm converges almost surely to an approximate Markov perfect equilibrium of the security game if the learning rates ${\alpha_k}$ satisfy the standard Robbins-Monro conditions.
To show convergence, we adopt an ordinary differential equation-based argument by the Borkar and Meyn theorem \cite{borkar1998stability}. We utilize properties of the queuing system and the feature functions to verify the Foster-Lyapunov drift condition \cite{meyn1993stability}. 
Instead of the typical requirement of uniform boundedness of feature functions, our result only requires the boundedness of the expectation with respect to the equilibrium distribution, which is much less restrictive. The result offers insights into the learning of effective and cost-aware defense mechanisms in real-life scenarios.

To assess the performance of the AMQ method, we conduct numerical experiments with a neural network Q (NNQ) function as the benchmark. We demonstrate that the AMQ method computes defense strategies that is $94.1\%-97.5\%$ (depending on the dimension of the approximators and the number of parallel servers) consistent with the NNQ method. The AMQ method approximates the equilibrium value function with an average error of $4.3\%-8.2\%$. Additionally, the NNQ method converges after approximately $2 \times 10^6$ iterations, whereas the AMQ method achieves convergence in around $10^4$ iterations. 
These results indicate that, in addition to the theoretical convergence guarantee, which is usually unavailable for neural networks, the AMQ method converges faster and attains an insignificant approximation error or optimality gap. 

In summary, our main contributions include:
\begin{enumerate}
  \item Development of an approximate minimax-Q learning algorithm for the Markov security game on parallel server systems,
  \item Convergence analysis of the proposed algorithm under rather mild assumptions, and
  \item Numerical experiments to validate the accuracy and efficiency of the proposed algorithm. 
\end{enumerate}

The rest of this paper is structured as follows. Section \ref{sec_model} presents the cyber-physical model and the approximate minimax-Q learning algorithm. Section \ref{sec_convergence} studies the convergence property of learning algorithm. Section \ref{sec_numer} conducts a numerical validation. Section \ref{sec_conclude} gives conclusions. 

\section{Model and Algorithm}
\label{sec_model}
In this section, we model the parallel server system and the strategic players, formulate the Markov security game, develop the function approximation scheme, and present the approximate minimax-Q learning algorithm.

\subsection{System and players}
Consider the parallel server system in Fig.~\ref{fig_system}. Jobs arrive according to a Poisson process of rate $\lambda>0$ and go to one of the $m$ servers. The $i$th server has exponentially distributed service times with service rate $\mu_i>0$. Let $x(t)\in \mathbb{Z}^m_{\geq 0}$ be the vector of the number of jobs in the servers, either waiting or being served. In the absence of attacks, we assume that an incoming job is routed to the server with the shortest queue; ties are broken uniformly at random. We select this policy because of its intuitiveness and popularity in practice \cite{singh2018throughput}.


We characterize the security problem as a two-player zero-sum game between a defender and an attacker. An attacker is able to manipulate the routing decision for an incoming job. The attacking cost is $c_1>0$ per unit time. 
A defender can defend the routing decision for an incoming job, at a cost of $c_2>0$ per unit time.
These costs account for the resources that attacking/defending actions have to consume.
If a routing decision is attacked and is not defended, the job will go to the longest server, as the consequence of a misled decision. Otherwise, the job will join the shortest queue correctly. Ties are broken uniformly at random. See Fig.~\ref{fig_system}.

The action space for the attacker is $\{0,1\}$, where $a(t)=0$ (resp. $a(t)=1$) means ``not to attack'' (resp. ``to attack'') at time $t.$ The action space for the defender is also $\{0,1\}$, where $b(t)=0$ (resp. $b(t)=1$) means ``not to defend'' (resp. ``to defend'') at time $t.$
The instantaneous reward (resp. cost) for the attacker (resp. defender) at time $t$ is defined as
\begin{align}
    \rho\Big(x(t),a(t),b(t)\Big):= \Vert x(t) \Vert_{1}-c_1a(t)+ c_2b(t),
    \label{Reward}
\end{align}
where $\|\cdot\|_1$ is the 1-norm. The action-induced costs are included in the reward/cost function, since both players may be interested in maximizing the opponent's costs. Note that the above reward/cost function assumes that both the traffic state and the opponent's action are observable to both players.
This assumption is technologically reasonable in many scenarios.
We are aware that there exist more sophisticated information structures in the literature; we do not consider them in this paper, since our focus is the coupling between the security game and the traffic dynamics. We believe that our analysis provides a basis for study of more sophisticated models.

Note that the queuing cost term $\|x\|_1$ will grow unboundedly (regardless of the players' actions) if the traffic demand $\lambda$ exceeds the total capacity $\sum_k\mu_k$. To exclude this less interesting case, we assume the following:
\begin{asm}\label{asm_lambda}
    The parallel server system is stabilizable in the sense that $\lambda<\sum_{n=1}^m \mu_n.$
\end{asm}
\noindent Under this assumption, the default join-the-shortest-queue routing policy is guaranteed to stabilize the traffic states if every job is routed correctly \cite{xie2024cost}. Hence, there exists at least one defending policy (i.e., $b(t)=1$ for all $t$) that ensures traffic stability.


\subsection{Security game}
Since we consider a version of off-policy learning algorithm, we differentiate the notations for the behavior policy and for the target policy.
We use $\alpha(a|x):\{0,1\}\times\mathbb{Z}_{\geq0}^m\to [0,1]$ (resp. $\beta(b|x):\{0,1\}\times\mathbb{Z}_{\geq0}^m\to[0,1]$) to denote the probabilistic behavior policy for the attacker (resp. defender).
We use $\pi(a|x):\{0,1\}\times\mathbb{Z}_{\geq0}^m\to [0,1]$ (resp. $\sigma(b|x):\{0,1\}\times\mathbb{Z}_{\geq0}^m\to [0,1]$) to denote the probabilistic target policy for the attacker (resp. defender).
Given a policy pair $(\alpha,\beta)$, the transition dynamics of the parallel server system can be specified as follows.
Let $e_i\in \{0,1\}^m$ denote the unit vector that has a 1 in the $i$-th entry and 0 elsewhere. 
Then the transition rate $q_{\alpha,\beta}:\mathbb Z_{\ge0}^m\times\mathbb Z_{\ge0}^m\to\mathbb R_{\ge0}$ of the traffic state under the policy pair $(\alpha,\beta)$ is given by
\begin{equation}
\begin{aligned}
    \label{transition_rate}
    q&_{\alpha,\beta}(y|x)= \\
   & \begin{cases}
 \Big(\frac{\alpha(0|x)}{|\arg\min_j x_j|}+\frac{\alpha(1|x)\beta(1|x)}{|\arg \min_j x_j|} \Big)\lambda  \\ \hspace{3.3cm} \mbox{if }y\in \{x+e_i;i\in\arg\min_jx_j\},\\
 \frac{\alpha(1|x)\beta(0|x)}{|\arg \min_j x_j|}\lambda \hspace{1.45cm} \mbox{if }y\in \{x+e_i;i\in\arg\max_jx_j\},\\
 \mu_i  \hspace{2.97cm} \mbox{if }y=x-e_i, \\
 0\hspace{3.1cm} \mbox{otherwise},
    \end{cases}
\end{aligned}
\end{equation}
where $|\cdot|$ denotes the cardinality of a set. We exclude the case of self-transition since it does not affect our analysis.


Since the system state is countable and changes only at discrete epochs, we can reformulate the Markov security game in discrete time (DT).
Note that a DT formulation also facilitates the design of learning algorithm.
Specifically, let $t_k$ be the $k$th transition epoch of the continuous-tim (CT) process $\{x(t);t\ge0\}$. With a slight abuse of notation, let
\begin{align*}
    x_k=x(t_k),\ 
    a_k=a(t_k),\ 
    b_k=b(t_k),\quad k=0,1,...
\end{align*}
Thus, the transition probabilities $p(x'|x,a,b)$ for the DT process can be obtained from the transition rates in \eqref{transition_rate} by the classical theory of countable-state Markov processes \cite{xie2024cost}. 
The expected one-step reward/cost is given by
\begin{equation}
    \begin{aligned}[b]
r\Big(x_k,&a_k,b_k\Big):=\\
&\rho\Big(x(t_{k-1}),a(t_{k-1}),b(t_{k-1})\Big)
 \mathbb E[\Delta t_k|x_k,a_k,b_k],
     \label{reward}
    \end{aligned}
\end{equation}
where $\Delta t_k=t_k-t_{k-1}$ is the exponentially distributed inter-transition interval.
The queuing dynamics ensure that $\mathbb E[\Delta t_k]$ exists for any $x_k,a_k,b_k$. Now we are ready to formally define the security game to be considered:

\begin{dfn}
    We consider a Markov game specified by a tuple $(\mathbb Z_{\ge0}^m,\mathcal{A},\mathcal{B},p,r,\gamma)$ defined as follows.
    \begin{enumerate}
        \item $\mathbb Z_{\ge0}^m$ is the traffic state space of the parallel server system.
        \item $\mathcal{A}$ (resp. $\mathcal{B}$) is the space of (mixed) strategies for the attacker (resp. defender).
        \item $p:(\mathbb Z_{\ge0}^m\times\{0,1\}^2)\times\mathbb Z_{\ge0}^m\to[0,1]$ is the transition probability of the traffic state under a given pair of actions; these probabilities can be computed readily from the CT transition rates $q$.
        \item $r:\mathbb Z_{\ge0}^m\times\{0,1\}^2\to\mathbb R$ is the one-step reward/cost function.
        \item $\gamma\in(0,1)$ is the discount rate.
    \end{enumerate}
\end{dfn}

By the DT formulation, the value/cost function is thus given by
$$
v_{\pi,\sigma}(x)=\mathbb E_{\pi,\sigma}\left[\sum\limits_{k=0}^{\infty}{\gamma^k r(x_k,a_k,b_k)\Bigg|x_0=x}\right].
$$
In the zero-sum game, the attacker (resp. defender) attempts to maximize (resp. minimize) the above. 
Since the state space is unbounded, the existence of $v_{\pi,\sigma}$ is not readily guaranteed for any policy pair. Fortunately, the existence was proved in \cite{xie2024cost} under equilibria in the following sense.
\begin{dfn}
    The Markov perfect equilibrium (MPE) for the security game is a strategy pair $(\pi^*,\sigma^*)$ such that for any $x \in \mathbb Z_{\ge0}^m$,
\begin{align*}
   &\pi^*(\cdot |x) = \arg \max_{\pi}v_{\pi,\sigma^*}(x),\\
   &\sigma^*(\cdot |x) = \arg \min_{\sigma}v_{\pi^*,\sigma}(x).
\end{align*}
\end{dfn}
Hence, the MPE is characterized by the equilibrium state value function
$$
v^*(x)=v_{\pi^*,\sigma^*}(x).
$$
Note that the corresponding action value function is given by
$$
Q_{\pi,\sigma}(x,a,b)
=r(x,a,b)+\sum_{x'\in \mathbb Z_{\ge0}^m}p(x'|x,a,b)v_{\pi,\sigma}(x').
$$
By the Shapley theory \cite{shapley1953stochastic}, $v^*$ is associated with a unique action value function (also called the ``minimax $Q$ function'') satisfying the minimax version of the Bellman equation \cite{szepesvari1999unified}.
Following \cite{zhu2020online}, we take the defender's perspective of minimax Bellman operator $\mathbf{T}$ on the space of functions $\{Q:\mathbb{Z}^m_{\geq 0}\times \{0,1\}^2 \rightarrow \mathbb{R} \}$ as
\begin{align*}
    (\mathbf{T}Q)&(x,a,b)= r(x,a,b)\nonumber\\
+&\gamma\min_{\sigma\in\mathcal B}\max_{a'\in\{0,1\}}\sum_{\substack{x'\in \mathbb Z_{\ge0}^m\\b'\in \{0,1\}}}p(x'|x,a',b')\sigma(b'|x)Q(x',a',b').
\end{align*}
Then the minimax Bellman equation can be written compactly as
\begin{align*}
    Q^*=\mathbf{T}Q^*,
\end{align*}
where $Q^*$ is also the action value function associated with $v^*.$

\subsection{Function Approximation}
Consider a set of $md$ linearly independent basis functions $\{\phi_{i,j}:\mathbb{Z}^m_{\geq 0}\times\{0,1\}^2 \rightarrow \mathbb{R};1\le i\le m,1\le j\le d\}$. 
Let
$$
\phi=[\phi_{1,1},\ldots,\phi_{1,d},\phi_{2,1},\ldots,\phi_{2,d},\ldots,\phi_{m,1},\ldots,\phi_{m,d}]^\top
$$
be the $md$-dimensional list of basis functions. 
We follow \cite{tanwani2015stability} and assume the following on regularity of the basis functions.
\begin{asm}\label{asm_featurefunc} 
    The basis functions $\phi$ satisfy the following. 
    \begin{enumerate}
    \item (Subexponential and non-negative) $\phi$ is such that
    \begin{align*}
       0\le \sum_{j=1}^d \phi_{i,j}(x)\leq e^{x_i} \ \ \ for\ i \in \{1,2,\dots,m\} .
    \end{align*}
    \item (Dominance over gradient) There exists a constant $B>0$ such that for $x$ satisfying
    \begin{align*}
        \|x\|_2^2\geq B,
    \end{align*}
    it holds that
    \begin{align*}
        \left\|\frac{\partial\phi}{\partial x}(x)\right\|_1<\|\phi(x)\|_1.
    \end{align*}
     \end{enumerate}
\end{asm}
\noindent Let 
$$\mathcal{Q}=\left\{\phi^\top w;\  w\in\mathbb{R}^{md}\right\}$$
be the space spanned by the basis functions.
Then the approximate function $Q_w\in \mathcal{Q}$ is given by
\begin{equation}
\label{approximationofQ}
    \begin{aligned}
    Q_{w}(x,a,b) = {\phi}(x,a,b)^\top w,
\end{aligned}
\end{equation}
where $w\in \mathbb{R}^{md}$ is the weight vector, with $w_{i,j}$ being the weight of $\phi_{i,j}$. Note that Assumption~\ref{asm_featurefunc} makes $w_{i,1},w_{i,2},\ldots,w_{i,d}$ associated with server $i$; this construction incorporates particularly the parallel structure of server system and thus gives interpretability of the weights.

If the behavior policy pair $(\alpha,\beta) \in \mathcal{A}\times \mathcal{B}$ ensures ergodicity of the traffic state process $\{x(t);t\ge0\}$, let $\mu_{\alpha,\beta}$ be the invariant probability measure. We will discuss the qualifications for the behavior policy in the next subsection.
With the linear function approximation, we in fact approximates the actual equilibrium value function $Q^*$ with a projection $Q_{w^*}$ in $\mathcal{Q}$.
Denote the orthogonal projection operator by $\mathbf{P}$ on the space of functions $\{Q:\mathbb{Z}^m_{\geq 0}\times \{0,1\}^2 \rightarrow \mathbb{R} \}$, which is given by 
\begin{equation}
    \begin{aligned}
    (\mathbf{P}Q)(x,a,b)=\phi^\top(x,a,b)\Sigma^{-1}\mathbb E_{\mu_{\alpha,\beta}}\left[\phi(x,a,b)Q(x,a,b)\right],\end{aligned}
    \end{equation}
where $\mathbb E_{\mu_{\alpha,\beta}}$, with a slight abuse of notation, denotes the vector of expectations with respect to the invariant probability measure $\mu_{\alpha,\beta}$.
The most intuitive approximation scheme is to directly project $Q^*$ on $\mathcal{Q}$ and obtain function $Q_{w^*}$ as
\begin{align*}
    Q_{w^*}(x,a,b)=(\mathbf{P}Q^*)(x,a,b)=(\mathbf{P}\mathbf{T}Q^*)(x,a,b).
\end{align*}
However, we generally can not obtain $Q^*$ analytically; otherwise we would not have to approximate. $Q_{w^*}$ here is also not a fixed point of any involved operator, and there exists no obvious procedure to write a stochastic approximation algorithm to find $Q_{w^*}$ \cite{melo2007convergence}.
Alternatively, we define the optimal weight vector $w^*$ to verify
\begin{align}
    Q_{w^*}(x,a,b)=(\mathbf{P}\mathbf{T}Q_{w^*})(x,a,b),
        \label{eq_PTQ}
    \end{align}
and approximate $Q^*$ with $Q_{w^*}$ as defined above. This $Q_{w^*}$ is actually a fixed point of the projected Bellman operator $\mathbf{P}\mathbf{T}$. Note that the corresponding optimal weight vector $w^*$ can also be directly defined as a fixed point of a modified projected Bellman operator.
Accordingly, we follow van Eck and van Wezel \cite{van2008application} and consider an approximated equilibrium as defined below:
\begin{dfn}
    The \emph{linear approximated equilibrium} for the security game is a strategy pair $(\hat\pi^*,\hat\sigma^*)$ such that for any $x \in \mathbb Z_{\ge0}^m$,
\begin{align*}
   &\hat\pi^*(\cdot |x)\\
   &= \arg \max_{\hat\pi\in\mathcal A}\sum_{a\in \{0,1\}}\hat\pi(a|x)\sum_{b\in \{0,1\}}\hat\sigma^*(b |x)
   \mathnormal{\phi}^\top  (x,a,b )w^*,\\
   &\hat\sigma^*(\cdot |x)\\
   &= \arg \min_{\hat\sigma\in\mathcal B}\sum_{b\in \{0,1\}}\hat\sigma(b|x)\sum_{a\in\{0,1\}}\hat\pi^*(a |x)\mathnormal{\phi}^\top (x,a,b)w^*.
\end{align*}
\end{dfn}
There are multiple metrics for the quality of approximation. One is the mean error between the actual value $Q^*$ and the approximated value $Q_{w^*}$. Another is the consistency between the MPE strategy profile $(\pi^*,\sigma^*)$ and the approximated MPE strategy profile $(\hat \pi^*,\hat \sigma^*)$. We will study these metrics numerically in Section~\ref{sec_numer}.

\subsection{Learning Algorithm}

We consider an approximate minimax-Q (AMQ) learning algorithm with the following update rule for the weights:
\begin{equation}
    \label{updaterule}
    \begin{aligned}[b]
        w_{k+1} &=w_k+\eta_k\nabla_w Q_w(x_k,a_k,b_k)\Delta_k \\ 
&=w_{k}+\eta_k\phi(x_{k},a_{k},b_{k})\Delta_{k}, 
    \end{aligned}
\end{equation}
where $\Delta_k$ is the temporal difference at time $t_k$, given by
\begin{equation}
    \begin{aligned}
\label{temporal_difference}
\Delta_k=&r_k+\gamma\min_{\sigma\in\mathcal{B}}\max_{a\in \{0,1\}}\sum_{b\in\{0,1\}}\sigma(b|x)Q_{w_k}(x_{k+1},a,b)\\ &-Q_{w_k}(x_k,a_k,b_k).
\end{aligned}
\end{equation}
To obtain $\sigma$ at iteration $k$, we actually solve a linear programming as follows, where the optimum objective $c=\max_{a\in \{0,1\}}\sum_{b\in\{0,1\}}\sigma(b|x)Q_{w_k}(x_{k+1},a,b)$. 
\begin{equation}
    \begin{aligned}
    \label{lp}
        \min & \quad \ c \\
        s.t  &\quad \sum_{b} \sigma(b|x)Q_{w_k}(x_{k+1},a,b) \leq c \qquad \ \forall a\in  \{0,1\}\\
        &\quad \ \sigma(b|x) \geq 0 ,\quad
        \quad \sum_{b} \sigma(b|x)=1 \qquad  \forall b\in \{0,1\}
    \end{aligned}
\end{equation}
The initial weight $w_0$ is arbitrary. The pseudo-code is presented below.

\begin{algorithm}[htb] 
\caption{AMQ learning for the security game}  
\begin{algorithmic}[1] 
\REQUIRE ~~\\ 
Initial weights $w_0$, behavior policy $\alpha,\beta$, step sizes sequence $\eta_k$, $\gamma$;\\
\STATE Initialize weights $w_0\leftarrow w_0$
    \FOR{$k=0,1,\cdots$}
        \STATE Sample $A_k\sim \alpha(\cdot|X_k)$, $B_k\sim \beta(\cdot|X_k)$
        \STATE Receive $R_{k+1}$ and observe $X_{k+1}$
        \STATE Update $\Delta_k$ via (\ref{temporal_difference}) and (\ref{lp})
        \STATE Update $w_k$ via (\ref{updaterule})
    \ENDFOR
 \end{algorithmic}
\end{algorithm}
We assume the following conditions for the behavior policy pair and for the learning rates.
\begin{asm}\label{behavior_policy}Let $(\alpha,\beta)\in\mathcal A\times\mathcal B$ be the behavior policy pair.
\begin{enumerate}
    \item $\alpha(a|x)> 0,\beta(b|x)> 0$ for $\mu_{\alpha,\beta}$-almost all $x\in \mathbb Z_{\ge0}^m$.
    \item There exist $\nu>0, c>0, d<\infty$ such that with $V(x)=\sum_{n=1}^{m}e^{\nu x_n}$,
\begin{align*}
    \mathcal{L}_{\alpha,\beta}V(x)&=\sum_{y\in \mathbb{Z}^m_{\geq 0}}q_{\alpha,\beta}(y|x)V(y)-V(x)\\ 
    &\leq -cV(x)+d,\qquad  \forall x\in\mathbb{Z}^m_{\geq 0},
\end{align*}
where $\mathcal{L}_{\alpha,\beta}$ is the infinitesimal generator under policy pair $(\alpha,\beta)$ and $q_{\alpha,\beta}(y|x)$ is defined in \eqref{transition_rate}. 
\end{enumerate}
\end{asm}
\begin{asm}\label{asm_alpha}
    The learning rates satisfy
    \begin{align*}
   \sum_{k=1}^\infty \eta_k = \infty, \quad \sum_{k=1}^\infty \eta_k^2 < \infty.
\end{align*}
\end{asm}
Assumption \ref{behavior_policy} ensures ergodicity under the behavior policy pair.
The class of policy pairs satisfying this assumption is fairly broad. In fact, any $\epsilon$-greedy-type policy pair would verify part 1). Part 2) essentially ensures positive Harris of the traffic state process. An illustrative example is provided below.
Under Assumption~\ref{asm_lambda}, there exists a positive constant $C_0$ satisfying $0< C_0<\min \{ 1,  \frac{\sum_{k=1}^m \mu_k-\lambda}{\lambda}\}$. Then a qualified behavior policy pair is
\begin{subequations}
\begin{align}
\label{e1}
&\alpha(1|x)=C_0e^{-\frac{|x|_1}{2}}, \ \ \ \alpha(0|x)=1-C_0e^{-\frac{|x|_1}{2}},\\
    \label{alpha}
&\beta(1|x)=
    \begin{cases}
 1-e^{-\frac{|x|_1}{2}} & \mbox{if }x\neq0^m,\\
 0.5 & \mbox{if }x=0^m.
    \end{cases}\\ 
    &\beta(0|x)=
    \begin{cases}
 e^{-\frac{|x|_1}{2}} & \mbox{if }x\neq0^m,\\
 0.5 & \mbox {if }x=0^m.
 \label{e3}
    \end{cases}
\end{align}
\end{subequations}
One can verify that this behavior policy pair satisfies Assumption~\ref{behavior_policy}; see Appendix. 
The assumptions on the learning rates are in fact the standard Robbins-Monro conditions for convergence analysis.

Finally, the AMQ learning algorithm is said to be convergent if $w_k\rightarrow w^*$ w.p.1, where $w^*$ verifies the projected Bellman equation \eqref{eq_PTQ}. The next section is devoted to show this.
\section{Convergence Analysis}
\label{sec_convergence}
The main result of this paper states that the approximate minimax-Q (AMQ) learning algorithm is guaranteed to converge to a solution to the projected minimax Bellman equation.

\begin{thm}
    Consider the Markov security game $(\mathbb Z_{\ge0}^m,\mathcal{A},\mathcal{B},p,r,\gamma)$ on a parallel server system. Under Assumptions~\ref{asm_lambda}--\ref{asm_alpha}, for any initial weight $w_0 \in \mathbb{R}^d$ and any initial state $x_0 \in \mathbb{Z}^m_{\geq 0}$, the approximate minimax-Q learning algorithm \eqref{updaterule} 
    converges in the sense that $w_k\rightarrow w^*$ w.p.1., where $w^*$ verifies the projected Bellman equation \eqref{eq_PTQ}.
    \label{theorem}
\end{thm}
Theorem \ref{theorem} provides a convergence guarantee for the proposed learning method under rather mild assumptions, viz. (i) stabilizability of the parallel server system, (ii) regularity of the basis functions, (iii) ergodicity under the behavior policy pair, and (iv) Robbins-Monroe conditions for the learning rates. Thus, the AMQ algorithm will reliably generate effective defense policies for managing parallel server systems in practical scenarios.

We will prove Theorem \ref{theorem} in three steps. In Section~\ref{sub_geometric}, we show that the traffic state is geometrically ergodic under the behavior policy pair (Lemma~\ref{lemma1}) and that the basis function has a bounded first moment with respect to the corresponding invariant probability measure (Lemma~\ref{lemma2}). In Section~\ref{sub_gradient}, we show that the first moment of the temporal-difference (TD) error is bounded by a linear function of the norm of the weight vector (Lemma~\ref{lemma3}).
In Section~\ref{proofoftheorem1}, we apply the ordinary differential equation-based argument to the first moment of the TD error and establish the convergence of the proposed algorithm.

\subsection{Geometric ergodicity and boundedness of basis functions}\label{sub_geometric}

Under a behavior policy pair $(\alpha,\beta)$ satisfying Assumption \ref{behavior_policy}, the induced chain $(\mathcal{X}, P_{\alpha,\beta})$ is geometrically ergodic with corresponding equilibrium probability measure $\mu_{\alpha,\beta}$. To argue for the irreducibility of the induced chain, note that
the state $x=0$ can be accessible from any initial condition with positive probability.
Hence, the induced chain is exponentially ergodic.

To prove the boundedness of feature functions, we first derive Lemma \ref{lemma1} to show the quadratic version of Lyapunov function $V(x)=\sum_{n=1}^{m}e^{\nu x_n}$ has a negative drift with $v>0$, based on which we can then conclude the boundedness of feature functions in Lemma \ref{lemma2}. 
\begin{lmm}
Suppose that assumption $1,3$ hold. Let $W(x)=(\sum_{n=1}^{m}e^{\nu x_n})^2$, $\nu>0$. Then there exist some $d'<\infty$ such that 
    \begin{align}
    &\mathcal{L}_{\alpha,\beta}W(x)=\sum_{y\in \mathbb{Z}^m_{\geq 0}}q_{\alpha,\beta}(y|x)W(y)-W(x)\leq -cW(x)+d', \nonumber\\
    &\hspace{6cm} x\in\mathbb{Z}^m_{\geq 0},
\end{align}
where $\mathcal{L}_{\alpha,\beta}$, $q_{\alpha,\beta}(y|x)$ and constant $c$ are defined in Assumption \ref{behavior_policy}.
\label{lemma1}
\end{lmm}
\noindent\textit{Proof.}
By Assumption \ref{behavior_policy} we obtain that 
\begin{align*}
\mathcal{L}_{\alpha,\beta}\Big(\sum_{n=1}^{m}e^{\nu x_n}\Big)\leq -c\Big(\sum_{n=1}^{m}e^{\nu x_n}\Big)+d, \quad x\in\mathbb{Z}^m_{\geq 0},
\end{align*}
where $c,d$ is finite constant defined in Assumption \ref{behavior_policy}. Note that $c>0$. 
Then the infinitesimal generator of $W(x)$ 
\begin{align*}
    \mathcal{L}_{\alpha,\beta}W(x) &= 2\Big(\sum_{n=1}^{m}e^{\nu x_n}\Big)\mathcal{L}\Big(\sum_{n=1}^{m}e^{\nu x_n}\Big) \\ &\leq -2c\Big(\sum_{n=1}^{m}e^{\nu x_n}\Big)^2+2d\Big(\sum_{n=1}^{m}e^{\nu x_n}\Big) \\
    &=-cW(x)+d',
\end{align*}
where $d'$ is a finite positive constant satisfying
\begin{align*}
    d' = -c\Big(\sum_{n=1}^{m}e^{\nu x_n}\Big)^2+2d\Big(\sum_{n=1}^{m}e^{\nu x_n}\Big) \leq \frac{d^2}{c}.
\end{align*}
$\qquad \qquad \qquad \qquad \qquad \qquad \qquad \qquad \qquad \qquad  \qquad \quad \ \ \ \ \qedsymbol$

Let $\Phi$ be the matrix defined as
$$\Phi=\mathbb E_{\mu_{\alpha,\beta}}\begin{bmatrix}\phi(x,a,b)\phi^\top(x,a,b)\end{bmatrix},$$
where $\mathbb E_{\mu_{\alpha,\beta}}$, with a slight abuse of notation, denotes the matrix of expectations with respect to the invariant probability measure $\mu_{\alpha,\beta}$.
The following result ensures the existence of $\Phi$.
\begin{lmm}
\label{lemma2}
Suppose that assumption $1-3$ hold, the feature function $\phi$ satisfies \begin{equation}
    \label{condition2}
    \begin{aligned}
    \Big\|\mathbb E_{\mu_{\alpha,\beta}}\Big[ \phi(x,a,b)\phi^\top(x,a,b,y)  \Big\|_{\infty}
   \leq \frac{d'}{c},
    \end{aligned}
\end{equation}
for any $i\in \{1,\ldots,m\}$, where $c,d'$ is the constant in Lemma $\ref{lemma1}$.
 
\end{lmm}
\noindent\textit{Proof.}
By Lemma \ref{lemma1} we obtain that
\begin{align*}
\lim_{t\to\infty}\frac1t\int_{s=0}^t \mathbb E \Big[\sum_{i=1}^{m}\sum_{j=1}^{m}e^{\nu\big(x_i(s)+x_j(s)\big)}\Big]ds\leq \frac{d'}{c} < \infty.
\end{align*}
Hence, by Assumption \ref{asm_featurefunc}, since
\begin{align*}
    \sum_{j=1}^d \phi_{i,j}(x)\leq e^{x_i} \ \ \ for\ i \in \{1,2,\dots,m\},
\end{align*}
then with $\psi(x)=\mathbb E_{\mu_{\alpha,\beta}}\big[\sum_{i=1}^{m}\sum_{j=1}^{d}\mathnormal{\phi}_{i,j}(x,a,b)\big]\leq\mathbb E_{\mu_{\alpha,\beta}}\big[\sum_{i=1}^{m} e^{x_i}\big]$ that
\begin{align*}
    \lim_{t\to\infty}\frac1t\int_{s=0}^t \mathbb E[\psi^2(x(s))]ds\leq \frac{d'}{c}
\end{align*}
for any initial condition $x(0)$. Then we can conclude that 
\begin{align*}
    \lim_{t\to\infty}\mathbb E[\psi^2(x(t))]\leq \frac{d'}{c},
\end{align*}
which means
\begin{align*}
    \Big\|\mathbb E_{\mu_{\alpha,\beta}}\Big[ \phi(x,a,b)\phi^\top(x,a,b,y) \Big\|_{\infty}
   \leq \frac{d'}{c},
\end{align*}
where $\mu_{\alpha,\beta}$ is the equilibrium state-action distribution under policy $\alpha,\beta$.
\qed

\subsection{Boundedness of gradient}\label{sub_gradient}
We write (\ref{updaterule}) in the form 
$$w_{k+1} =w_{k}+\eta_k H(w_k,Y_{k+1}),$$ 
where $Y_{k+1}=(x_k, a_k, b_k)$, and
\begin{equation}
    \begin{aligned}[b]
      H(w,Y)=&\phi(x,a,b)\Big(r(x,a,b,y)
+\\ \gamma \min_{\sigma\in\mathcal{B}}&\max_{a'\in \{0,1\}}\sum_{b'\in \{0,1\}} \sigma(b^{\prime})Q_{w}(y,a^{\prime},b^{\prime})-Q_{w}(x,a,b)\Big).
\label{Hfunc}
    \end{aligned}
\end{equation}
To prove the boundedness of the gradient, we mainly utilize the properties of feature function and queuing system. 

\begin{lmm}
\label{lemma3}
 The function $H$ satisfies \begin{equation}
    \label{condition3}
    \begin{aligned}
    \Big\|\mathbb E_{\mu_{\alpha,\beta}}\big[ H(w,x,a,b) \big ]\Big\|_{\infty}
     \leq C(1+\|w\|_{\infty}),
    \end{aligned}
\end{equation}
for any $w$, where $C$ is a finite constant.

\end{lmm}
\noindent\textit{Proof.}
Denote $e_i$ as the unit vector with only the $i$th element equals $1$. Also denote the longest queue as $x_{\max}$ and its corresponding index as $i$. Define similarly the shortest queue $x_{\min}$ and its index $j$.
Denote by $g(x,a,b,y)$ the vector as
\begin{align*}
    g(x,a,b,y) =\max_{\substack{a^{\prime} \in\{0,1\}\\b^{\prime}\in\{0,1\}}} \phi(y,a^{\prime},b^{\prime})-\phi(x,a,b)
\end{align*}
Hence, we can obtain by definition of (\ref{Hfunc}) that
\begin{equation}
\begin{aligned}
    &\Big\|\mathbb E_{\mu_{\alpha,\beta}}\big[ H(w,x,a,b) \big ]\Big\|_{\infty}\leq \\ &\leq \Big\|\mathbb E_{\mu_{\alpha,\beta}}\Big[\phi(x,a,b)\big[r(x,a,b,y)+g^\top(x,a,b,y)\cdot w \big] \Big] 
 \Big\|_{\infty}  \\ 
    &\leq \Big\|\mathbb E_{\mu_{\alpha,\beta}}\big[\phi(x,a,b) \cdot r(x,a,b,y)\big]\Big\|_{\infty} \\& \qquad +\Big\|\mathbb E_{\mu_{\alpha,\beta}}\big[\phi(x,a,b)\cdot g^\top(x,a,b,y)\big]\Big\|_{\infty}\cdot \|w\|_{\infty}
    \label{lmm3_detail}
\end{aligned}
\end{equation}


When $\sum_{k=1}^{m}x_k^2\geq B$, it can be deduced that
$ \|g^\top(x,a,b,y)\|_{1}\leq \|\phi^\top(x,a,b)\|_{1}$ by Assumption \ref{asm_featurefunc}. Thus by Lemma $\ref{lemma2}$, 
\begin{align*}
    &\Big\|\mathbb E_{\mu_{\alpha,\beta}}\Big[ \phi(x,a,b)g^\top(x,a,b,y)\Big|\sum_{k=1}^{m}x_k^2\geq B\Big] \\& \cdot P\Big(\sum_{k=1}^{m}x_k^2\geq B \Big) \Big\|_{\infty}\\& + 
    \Big\|\mathbb E_{\mu_{\alpha,\beta}}\Big[ \phi(x,a,b)g^\top(x,a,b,y)\Big|\sum_{k=1}^{m}x_k^2< B\Big] \\& \cdot P\Big(\sum_{k=1}^{m}x_k^2< B \Big) \Big\|_{\infty} \\&< \frac{d'}{c}+ \Big\|\mathbb E_{\mu_{\alpha,\beta}}[\phi(x,a,b)(\zeta(B))^2 ]\Big\|_{\infty}
\end{align*}
where $c,d'$ is the constant in Lemma $\ref{lemma1}$, $\zeta(B)$ is a finite positive constant related to the specific form of $\phi$. It can be easily derived according to different combination of state action pairs, e.g. $\zeta(B)=(\sqrt{B}+1)^2$ when adopting polynomial approximators. By Assumption \ref{behavior_policy}, we obtain
\begin{align*}
\lim_{t\to\infty}\frac1t\int_{s=0}^t \mathbb E \Big[\sum_{i=1}^{m}e^{\nu\big(x_i(s)\big)} \Big]ds\leq \frac{d}{c} < \infty.
\end{align*}
where $d$ is the constant in Assumption \ref{behavior_policy}. Hence, we can derive with $\psi(x)=\mathbb E_{\mu_{\alpha,\beta}}\big[\|\phi(x,a,b)\|_{1}\big],$
\begin{align*}
    \lim_{t\to\infty}\frac1t\int_{s=0}^t\mathbb E[\psi(x(s))]ds\leq \frac{d}{c}
\end{align*}
for any initial condition $x(0)$. Then we conclude that
\begin{align*}
    \lim_{t\to\infty}\mathbb E[\psi(x(t))]\leq \frac{d}{c},
\end{align*}


which implies
\begin{align*}
    \Big\|\mathbb E_{\mu_{\alpha,\beta}}[\phi(x,a,b) ]\Big\|_{\infty} \leq \frac{d}{c}.
\end{align*}
Hence, 
\begin{equation}
    \begin{aligned}
   \Big \|\mathbb E_{\mu_{\alpha,\beta}}[ \phi(x,a,b)g^\top(x,a,b,y)] \Big\|_{\infty} < \frac{1}{c}\Big(d'+d(\zeta(B))^2\Big).
    \label{lmm3_condition}
\end{aligned}
\end{equation}

Then we prove the term $\|\mathbb E_{\mu_{\alpha,\beta}}\big[\phi(x,a,b) \cdot r(x,a,b,y)\big]\|_{\infty}$ is bounded.
Recall the definition of reward (\ref{reward}). We first denote the interval time until next arrival as $\Delta t^1$ with its distribution $f_a(\Delta t^1)$, the interval time until next service as $\Delta t^2$ with its distribution $f_s(\Delta t^2)$. The number of current activated servers (i.e. with job in queue) is defined as
\begin{align*}
    k_x:=k(x)=\sum_{i=0}^{m} \mathbb{I} \{x_i \neq 0 \}.
\end{align*}
It is known from property of parallel queuing system that
\begin{align*}
    f_a(\Delta t^1) &= \lambda \exp{(-\lambda \Delta t^1)}\\
    f_s(\Delta t^2) &= k_x\mu \exp{(-k_x\mu \Delta t^2)}.
\end{align*}
Since $\Delta t=\min \{ \Delta t^1, \Delta t^2 \}$, we can derived the distribution of $\Delta t$ as $f(\Delta t)$ 
\begin{align*}
    f(\Delta t)=(\lambda + k_x\mu) \exp (-(\lambda+k_x\mu)\Delta t),
\end{align*}
with expectation of $\Delta t$ as $\mathbb E[\Delta t]=\frac{1}{\lambda+k_x\mu} \leq \frac{1}{\lambda}$.
By definition of (\ref{Reward}) and the fact our state $x\in \mathbb{Z}^m_{\geq 0}$,
\begin{align*}
    \rho(x,a,b)\leq \|\phi^\top(x,a,b)\|_{1}.
\end{align*}
Then by Lemma \ref{lemma2}, we can conclude
\begin{align*}
    &\Big\|\mathbb E_{\mu_{\alpha,\beta}}\big[\phi(x,a,b) \cdot r(x,a,b,y)\big]\Big\|_{\infty}
    \\&\leq \frac{1}{\lambda}\Big\|\mathbb E_{\mu_{\alpha,\beta}}\Big[ \phi(x,a,b)\phi^\top(x,a,b,y)  \Big\|_{\infty}
    \leq \frac{d'}{c\lambda}.
\end{align*}
Hence, $(\ref{condition3})$ can be satisfied by selecting $C = \max \Big\{\frac{1}{c}\Big(d'+d(\zeta(B))^2\Big),\frac{d'}{c\lambda}\Big\}$.$\qquad \qquad \qquad \qquad \qquad  \qquad \qquad \qquad \ \ \qedsymbol$

\subsection{Proof of Theorem 1}
\label{proofoftheorem1}
We first prove the convergence of approximate minimax Q learning w.p.1. Let $\mu_x$ be the corresponding invariant probability measure, $\mu_{\alpha,\beta}$ be the invariant state-action distribution under the given behavior policy pair $(\alpha, \beta)$. It verifies the existence of function
\begin{align*}
    h(w) = \int H(w,Y)\mu_{\alpha,\beta}(dY)
\end{align*}
by bound of function $H(w,Y)$ derived in Lemma $\ref{lemma3}$.

Since the chain is geometrically ergodic, 
it follows that so is the chain ${Y_k}$. The geometric ergodicity of ${Y_k}$ and the fact that $\alpha,\beta$ do not depend on $w$ ensure that the requirements are satisfied. Hence, by \cite[Theorem 17]{benveniste2012adaptive} , the convergence of ${w_k}$ w.p.1 is established as long as the ODE
\begin{equation}
    \label{ODE}
    \begin{aligned}
        \dot{w}_k=h(w_k)
    \end{aligned}
\end{equation}
with 
\begin{align*}
    h(w)&=\mathbb E_{\mu_{\alpha,\beta}} \Big [ \phi(x,a,b)\Big(r(x,a,b,y)+ \\  +\gamma & \min_{\sigma\in\mathcal{B}}\max_{a'\in\{0,1\}}\sum_{b'\in\{0,1\}} \sigma(b^{\prime}) \phi^\top (y,a^{\prime},b^{\prime})w-\phi^\top (x,a,b)w \Big) \Big ],
\end{align*}
has a globally asymptotically stable equilibrium $w^*$.

We can write $h$ as 
$$h(w)=h_1(w)-h_2(w),$$ 
with 
\begin{align*}
    h_1(w)=&\mathbb E_{\mu_{\alpha,\beta}}[\phi(x,a,b)(r(x,a,b,y)+ \\ &+\gamma \min_{\sigma\in\mathcal{B}}\max_{a'\in\{0,1\}}\sum_{b'\in\{0,1\}} \sigma(b^{\prime}) \phi^\top (y,a^{\prime},b^{\prime})w)]
\end{align*}
and
\begin{align*}
    h_2(w)=\mathbb E_{\mu_{\alpha,\beta}}[\phi(x,a,b)\phi^\top (x,a,b)w)].
\end{align*}
Then using the non-expansiveness of \textit{min} and \textit{max} operator, we can conclude
    \begin{align*}
        &\|h_1(w_1)-h_1(w_2)\|_{\infty}=\\
    &\|\mathbb E_{\mu_{\alpha,\beta}}[\gamma \phi(x,a,b) ( \min_{\sigma\in\mathcal{B}}\max_{a'\in\{0,1\}}\sum_{b'\in\{0,1\}} \sigma(b^{\prime}) \phi^{\top}(y,a^{\prime},b^{\prime})w_1 \\ &-\min_{\sigma\in\mathcal{B}} \max_{a'\in\{0,1\}}\sum_{b'\in\{0,1\}} \sigma(b^{\prime}) \phi^{\top}(y,a^{\prime},b^{\prime})w_2  ) ]\|_{\infty} \\ &\leq
    \|\mathbb E_{\mu_{\alpha,\beta}}[\gamma \phi(x,a,b)  \max_{\sigma\in\mathcal{B}}( \max_{a'\in\{0,1\}}\sum_{b'\in\{0,1\}} \sigma(b^{\prime}) \phi^{\top}(y,a^{\prime},b^{\prime})w_1 \\ &- \max_{a'\in\{0,1\}}\sum_{b'\in\{0,1\}} \sigma(b^{\prime}) \phi^{\top}(y,a^{\prime},b^{\prime})w_2  ) ]\|_{\infty} \\& \leq 
    \gamma \|\mathbb E_{\mu_{\alpha,\beta}}[ \phi(x,a,b) \max_{\sigma\in\mathcal{B}}\max_{a'\in\{0,1\}} \sum_{b'\in\{0,1\}} \sigma(b^{\prime}) \phi^{\top}(y,a^{\prime},b^{\prime})\\ & 
    (w_1-w_2)   ]\|_{\infty}  
    \\&\leq 
    \gamma \Big ( \|\mathbb E_{\mu_{\alpha,\beta}}[ \phi(x,a,b)\phi^{\top}(x,a,b)] \|_{\infty} + \\ &\|\mathbb E_{\mu_{\alpha,\beta}}[ \phi(x,a,b)g^{\top}(x,a,b,a',b')] \|_{\infty} \Big ) \cdot \|w_1-w_2\|_{\infty},
    \end{align*}
where $g(x,a,b,y)$ is defined in Lemma \ref{lemma3}.

Actually, we can scale the feature function $\phi(x,a,b)$ arbitrarily to make $h_1$ be $\gamma$-contraction. Scale $\phi(x,a,b)$ by a constant factor $\varepsilon\leq\frac{\sqrt{[d'+d(\zeta(B))^2]^2+4d'c}-[d'+d(\zeta(B))^2]}{2d'}$, where $B$ is the constant defined in Assumption $\ref{asm_featurefunc}$.
Then by Lemma \ref{lemma2} and condition (\ref{lmm3_condition}) in Lemma \ref{lemma3} we can ensure 
\begin{equation}
\begin{aligned}
\label{h1_contraction}
    \|h_1(&w_1)-h_1(w_2)\|_{\infty}= \quad \\\leq 
    \gamma \Big ( &\varepsilon^2\|\mathbb E_{\mu_{\alpha,\beta}}[ \phi(x,a,b)\phi^{\top}(x,a,b)] \|_{\infty} + \\ &\varepsilon\|\mathbb E_{\mu_{\alpha,\beta}}[ \phi(x,a,b)g^\top(x,a,b,y)] \|_{\infty} \Big ) \cdot \|w_1-w_2\|_{\infty} \\ \leq \gamma\Big[&\frac{\varepsilon^2d'}{c}+\varepsilon\Big(\frac{d'}{c}+\frac{d}{c}\big(\zeta(B))^2\Big)\Big]\cdot \|w_1-w_2\|_{\infty} \\\leq \gamma\|&w_1-w_2\|_{\infty}.
\end{aligned}
\end{equation}
Also, we can conclude by Lemma \ref{lemma2} that
\begin{equation}
    \label{h2contraction}
    \begin{aligned}
        \|h_2(w_1)-h_2(w_2)\|_{\infty}&=\\ 
        \|\mathbb E_{\mu_{\alpha,\beta}} [\phi(x,a,b)\phi^\top (x&,a,b)(w_{1}-w_{2})]\|_{\infty}   \leq
    \| (w_1-w_2)  \|_{\infty}.
    \end{aligned}
\end{equation}
Next we calculate the derivative of $p$-norm of term $(w_k-w^*)$, where $w^*$ is the equilibrium point of (\ref{ODE}) which verifies $h(w^*)=0$.
\begin{align*}
    \frac{d}{dk}\|w_{k}&-w^{*}\|_{p}= \|w_{k}-w^{*}\|_{p}^{1-p}\\
    \cdot \Big (&\sum_{i=1}^{d}(w_{k}(i)-w^{*}(i))^{p-1}\cdot\big((h_{1}(w_{k}))_{i}-(h_{1}(w^{*}))_{i}\big)+\\  &\sum_{i=1}^{d}(w_{k}(i)-w^{*}(i))^{p-1}\cdot\big((h_{2}(w^{*}))_{i}-(h_{2}(w_{k}))_{i}\big)\Big ),
\end{align*}
where we denote by $(h_1(w))_i$ the $i^{th}$ component of $h_1(w)$ and similarly for $h_2$. Applying Hölder’s inequality to the above summations yields
\begin{align*}
    \frac{d}{dk}\|w_{k}-w^{*}\|_{p}\leq \|h_{1}(w_k)-h_{1}(w^{*})\|_{p}+\|h_{2}(w^{*})-h_{2}(w_k)\|_{p}.
\end{align*}
Taking the limit as $p\rightarrow\infty $ and using (\ref{h1_contraction}) and (\ref{h2contraction}) leads to
\begin{equation}
    \label{derivative}
    \begin{aligned}
    \frac{d}{dk}\|w_{k}-w^{*}\|_{\infty}&\leq (\gamma-1)\|w_{k}-w^{*}\|_{\infty}.
\end{aligned}
\end{equation}
Let $\lambda=1-\gamma>0$. Integrate w.r.t $k$, (\ref{derivative}) becomes
\begin{align*}
    \|w_{k}-w^{*}\|_{\infty}\leq e^{-\lambda k}\|w_{0}-w^{*}\|_{\infty},
\end{align*}
which establishes the existence of a globally asymptotically stable equilibrium point for (\ref{ODE}).
And it is clear that $h(w^*)=0$
leads to
\begin{equation}
    \label{stablepoint}
    \begin{aligned}
    w^*=\Sigma^{-1} \mathbb E_{\mu_{\alpha,\beta}}[\phi(x,a,b)(r(x,a&,b,y)+ \\ \gamma \min_{\sigma\in\mathcal{B}}\max_{a'\in\{0,1\}}&\sum_{b'\in\{0,1\}} \sigma(b^{\prime}) \phi^\top (y,a^{\prime},b^{\prime})w^*)].
\end{aligned}
\end{equation}
Hence, the sequence ${w_k}$ converges w.p.1 to the globally asymptotically stable equilibrium point $w^*$.

Then we further prove that the limit of approximate minimax-Q function is the fixed point of projected Bellman operator. 
Given $w^*$ as (\ref{stablepoint}), the corresponding approximate $Q$ function
\begin{align*}
    Q_{w^{*}}(x,a,b)&=\\&\phi^{\top}(x,a,b)\Sigma^{-1}\mathbb E_{\mu_{\alpha,\beta}}\left[\phi(x,a,b)(\mathbf{T}Q_{w^{*}})(x,a,b)\right]\\&=(\mathbf{P} \mathbf{T}Q_{w^*})(x,a,b).
\end{align*}
This implies that $Q_{w^*}$ verifies the fixed point equation in (\ref{eq_PTQ}).
\section{Numerical Validation}
\label{sec_numer}

In this section, we implement the approximate minimax-Q (AMQ) learning algorithm and numerically evaluate its performance. The objectives of this section is (i) to present and interpret the cost-aware defending strategy given by the AMQ method and (ii) to study the computational efficiency and approximation accuracy of the AMQ method.

\subsection{Experiment setup}

We simulate two system models, one with three parallel servers and one with six; this is intended to study the impact of system complexity. The service rates are listed in Table~\ref{tab_para}:
\begin{table}[H]
	\centering
	\caption{Experiment parameters.}
	\label{tab_para}  
	\begin{tabular}{ccc}
		\hline\noalign{\smallskip}	
		Parameter & Notation & Value \\ 
		\noalign{\smallskip}\hline\noalign{\smallskip}
         Arrival rate & $\lambda$ & 5 per unit time\\ 
         Service rate 1 & $\mu_1$ & 2 per unit time \\
         Service rate 2 & $\mu_2$ & 3 per unit time \\
         Service rate 3 & $\mu_3$ & 4 per unit time \\
         Service rate 4 & $\mu_4$ & 2 per unit time \\
         Service rate 5 & $\mu_5$ & 0.5 per unit time \\
         Service rate 6 & $\mu_6$ & 1 per unit time \\
         Attacking cost & $c_1$ & 8 per unit time \\
         Defending cost & $c_2$ & 6 per unit time \\
         Discount factor & $\gamma$ & 0.9 \\
         Behavior policy constant & $C_0$ & 0.6\\
		\noalign{\smallskip}\hline
	\end{tabular}
\end{table}
\noindent$\mu_1$--$\mu_3$ are used for the three-server model, while $\mu_1$--$\mu_6$ are used for the six-server model. The table also gives the other parameters.
The policies given by \eqref{e1}--\eqref{e3} are used as the behavior policies.
The initial target policies are set to be the random policies $\sigma(0|x)=\sigma(1|x)=0.5$ and $\pi(0|x)=\pi(1|x)=0.5$ for all $x\in\mathbb Z_{\ge0}^m$.
The initial traffic state is randomly generated.


We use a neural network Q (NNQ) learning as the benchmark for evaluate the AMQ method.
The NNQ methods approximates the value function $Q(x,a,b)$ with a neural network and trains it according to the minimax Bellman equation. 
Since NNs have extremely strong approximation performance, we use the NNQ function as a proxy for the ground truth of the equilibrium value, which cannot be analytically obtained. The architecture of the NN comprises two fully connected layers, employing a rectified linear unit (ReLU) as the activation function. The NN is updated via adaptive moment estimation. The loss function used is the mean squared error between the predicted one-step and calculated state-action value. 

For the AMQ method, we consider two approximators with different dimensions. The first, named ``AMQ1'', is a collection of affine functions of the traffic states: for $i=1,2,\ldots,m$,
\begin{align*}
    &\phi_{i,1}(x,a,b)=1,\quad\phi_{i,2}(x,a,b)=x_i+\delta_i(x,a,b),\\
    &\phi_{i,3}(x,a,b)=a,\quad\phi_{i,4}(x,a,b)=b,
\end{align*}
where $\delta_i(x,a,b)$ is given by
\begin{align*}
\delta_i(a,b):=
    \begin{cases}
 1 & \mbox{if }i = \arg\max_{i} x_i,(a,b)=(1,0),\\
 1 & \mbox{if }i=  \arg\min_{i} x_i,(a,b)\neq (1,0), \\
 0 & \mbox{otherwise}.
    \end{cases}
\end{align*}
Intuitively, the feature functions are motivated by the reward function in \eqref{Reward}.
The second, named ``AMQ2'', is a collection of second-order polynomials of the traffic states: for $i=1,2,\ldots,m$,
\begin{align*}
    &\phi_{i,1}(x,a,b)=1,\quad\phi_{i,2}(x,a,b)=x_i+\delta_i(x,a,b),\\
    &\phi_{i,3}(x,a,b)=\Big(x_i+\delta_i(x,a,b)\Big)^2,\\
    &\phi_{i,4}(x,a,b)=a,\quad\phi_{i,5}(x,a,b)=b.
\end{align*}
Hence, AMQ2 is more flexible than AMQ1 and will turn out to be more accurate than AMQ1.

Table~\ref{algorithms} summarizes the three algorithms that we consider. Note that they are all off-policy temporal-difference learning methods.

\begin{table}[H]
	\centering
	\caption{Algorithms to be compared.}
	\label{algorithms}  
	\begin{tabular}{cc}
		\hline\noalign{\smallskip}	
		Algorithm & Approximator\\ 
		\noalign{\smallskip}\hline\noalign{\smallskip}
         NNQ (Baseline) & Two-layer neural network with ReLU\\ 
         AMQ1 (Ours) & Affine functions of traffic state\\
         AMQ2 (Ours) & Second-order polynomials of traffic state\\
		\noalign{\smallskip}\hline
	\end{tabular}
\end{table}

 We trained and evaluated the learning algorithms for $2\times10^6$ epochs. A discrete time step of 0.1 seconds was employed for simulation. All experiments were conducted using Jupyter Notebook, hosted on a system equipped with an Intel(R) Xeon(R) CPU with 36.7 GB of memory. 

\subsection{Interpretation of trained weights}

Every experiment that we conducted converged to an approximate equilibrium. As an illustration, consider the three-server setting. The weights for the AMQ2 in this setting turn out to be
\begin{center}
\begin{tabular}{lll}
     $w_{1,1}=6.55$,&$w_{2,1}=5.55$,&$w_{3,1}=4.55$,\\ 
    $w_{1,2}=9.74$,&$w_{2,2}=9.23$,&$w_{3,2}=9.02$,\\ 
    $w_{1,3}= 0.46$,&$w_{2,3}=0.41$,&$w_{3,3}=0.34$,\\ 
    $w_{1,4}= 0.9$,&$w_{2,4}=0.8$,&$w_{3,4}=0.8$,\\ 
    $w_{1,5}= -1.1$,&$w_{2,5}=-1.0$,&$w_{3,5}=-0.89$.
\end{tabular}
\end{center}
Recall that the function $\phi^\top w^*$ is the (approximate) equilibrium cost for the defender; the first index in the subscript is actually the server index.

There are several insights about the weights associated with the same server worth mentioning.
First, the first-order terms are associated with weights ($w_{i,3}$) greater than the second-order terms ($w_{i,2}$); this implies that the value function grows roughly linearly with the traffic states.
Second, a non-trivial intercept exists ($w_{i,1}$) for every server, which implies that a server might be associated with a risk even if it is idling. 
Finally, the weights ($w_{i,4},w_{i,5}$) associated with the player actions have the correct signs. In addition, attacks are associated with smaller weights than defenses, so the defender seems to have a stronger incentive to defend than the attacker to attack; this is probably due to that the defending cost is lower than the attacking cost.

Across various servers, it turns out that queues with lower service rates are in general associated with higher risks, which is intuitive.
Interestingly, the greater intercepts ($w_{i,1}$) are consistently associated with higher service rates; that is, an incorrect routing to a slow server, even if it is idling, may still be costly.
In addition, the weights ($w_{i,4},w_{i,5}$) associated with the player actions directly indicates the benefit of attacking/defending a particular server; servers with slower service rates are associated with, without surprise, higher weights.


\subsection{Evaluation of algorithm}
Table \ref{tablei} presents the normalized learned values and policies with respect to the equilibrium state distribution. The initial state is sampled from this equilibrium distribution, and empirical data is obtained using the Monte Carlo method. The reported results represent the average of 10 repeated experiments. The findings indicate that the learned results of AMQ2 approximate optimal defense strategies with an average error of $2.5\%$, and approximate the optimal values with an average error of $4.3\%$ under the equilibrium distribution, thus validating the precision of the proposed algorithm in approximating both optimal values and optimal policies. The performance of the AMQ2 algorithm further highlights that the inclusion of quadratic terms in the feature functions improves the empirical average cost by $3.6\%$ and the empirical policy consistency by $3.3\%$. These results underscore the necessity of incorporating quadratic feature functions to achieve more accurate learning outcomes.

\begin{table}[H]
	\centering
	\caption{Performance of various methods}
	\label{tablei}  
	\begin{tabular}{ccccc}
		\hline\noalign{\smallskip}	
		Metric & System  & AMQ1 & AMQ2&  NNQ\\ 
		\noalign{\smallskip}\hline\noalign{\smallskip}
         Normalized mean cost & 3-server  & 1.079 & 1.043& 1.000\\
         Policy consistency & 3-server  & 94.2\% & 97.5\%& 100\%\\
         \noalign{\smallskip}\hline\noalign{\smallskip}
        Normalized mean cost & 6-server  & 1.082 & 1.045& 1.000\\
         Policy consistency & 6-server  & 94.1\% & 97.3\%& 100\%\\		\noalign{\smallskip}\hline
	\end{tabular}
\end{table}

\begin{figure}[H]
  \centering
  \includegraphics[width=0.48\textwidth]{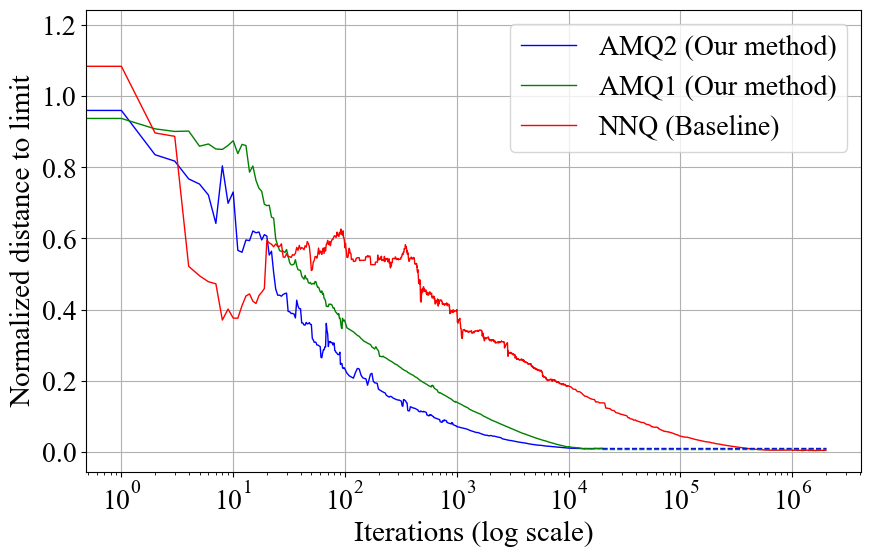}
  \caption{Performance comparison on distance to limit.}
  \label{fig_error}
\end{figure}

Fig.~\ref{fig_error} illustrates the normalized $l_2$-norm difference between the weights $w_t$ and the optimal weights $w^*$ throughout the learning process for both the three methods. It is evident that the NN method converges after approximately $2.4 \times 10^5$ iterations, whereas the AMQ1 and the AMQ2 method achieves convergence after around $5\times 10^3$ iterations. Hence, our proposed algorithm has a much higher convergence rate compared to NN, validating the efficiency of the AMQ learning algorithm.

To test the scalability of AMQ, we further implement identical experiments on six servers. The results are also shown in Table \ref{tablei}. It can be seen that in six servers setting, the performance of AMQ degrades at most $0.2\%$ in approximating optimal defense strategies and $0.3\%$ in approximating the optimal values, compared to the three servers case. 
The results indicate that the computational advantage of linear approximation remains at more servers.  
\section{Concluding remarks}
\label{sec_conclude}
This paper considers securing parallel server systems against malicious cyber-physical attacks. The proposed approximate minimax-Q (AMQ) learning algorithm efficiently balances security costs and performance losses. The algorithm uses an interpretable linear approximation scheme and adapts to the system’s structure. A key advantage of this method compared with deep reinforcement learning methods is a theoretical guarantee for convergence with probability one to an equilibrium under mild assumptions. We established this result by combining the stability theory of Markov processes and an ordinary differential equation-based technique. Tests show the AMQ learning converges faster than neural networks, with an insignificant optimality gap. The approach combines theory and practice, offering scalable security for cloud, manufacturing, and transport systems. It highlights reinforcement learning’s potential in adversarial settings with complex, unbounded state spaces.
Future work will extend the framework to payoff-based learning algorithms and partially observable environments. 
\begin{appendix}
\section{Appendices}
Lemma \ref{lmm_policyveri} shows that there exist behavior policies satisfying Assumption \ref{behavior_policy}. 
\begin{lmm}
\label{lmm_policyveri}
Under the policy pair (\ref{e1})-(\ref{e3}). Suppose that assumption $1$ holds. Let $V(x)=\sum_{n=1}^{m}e^{vx_n}$, $v>0$. Then there exist some $c>0, d<\infty$ such that 
\begin{align}
    &\mathcal{L}_{\alpha,\beta}V(x)=\sum_{y\in \mathbb{Z}^m_{\geq 0}}q_{\alpha,\beta}(y|x)V(y)-V(x)\leq -cV(x)+d, \nonumber\\ 
    &\hspace{6cm} x\in\mathbb{Z}^m_{\geq 0},\label{condition}
\end{align}
where $\mathcal{L}_{\alpha,\beta}$ is the infinitesimal generator under policy pair $\alpha,\beta$, $q_{\alpha,\beta}(y|x)$ are the transition rates from state $x$ to $y$ defined in (\ref{transition_rate}).
\end{lmm}
\noindent\textit{Proof.}
Denote the longest queue as $x_{\max}$ and its corresponding index as $i$. Define similarly the shortest queue $x_{\min}$ and its index $j$. Let $l_n=\mathbb{I}_{\{x_n\geq1\}}$.We have
\begin{align*}
\mathcal{L}_{\alpha,\beta}V(x)=& \sum_{n=1}^{m}l_n\mu_n(e^{v(x_n-1)}-e^{vx_n} )\\+&\lambda\Big(C_0e^{-|x|_1}\cdot (e^{v(x_{\max}+1)}-e^{x_{\max}})\\
+& (1-C_0e^{-|x|_1})\cdot (e^{v(x_{\min}+1)}-e^{x_{\min}})  \Big).
\end{align*}
Note that $l_n\cdot e^{vx_n}=(l_n-1)+e^{vx_n}$, so we have
\begin{equation}
\label{drift}
\begin{aligned}[b]
\mathcal{L}_{\alpha,\beta}&V(x)= \sum_{\substack{n=1\\n\neq i,j }}^{m}\mu_ne^{vx_n}(e^{-v}-1 ) + B_{0}\\
    +&\Big (\mu_i(e^{-v}-1)+\lambda(e^v-1)C_0e^{-|x|_1}\Big ) e^{vx_{\max}}\\
+& \Big (\mu_j(e^{-v}-1)+\lambda(e^v-1)(1-C_0e^{-|x|_1})\Big)e^{vx_{\min}}.
\end{aligned}
\end{equation}
where $B_0=\sum_{n=1}^{m} (l_n-1)\mu_n(e^{-v}-1)$ is a finite non-negative constant. It can be deduced that the drift equation (\ref{drift})
\begin{align*}
    \mathcal{L}_{\alpha,\beta}V(x)& < \sum_{\substack{n=1\\n\neq i,j }}^{m} e^{vx_n} \cdot f_n(v) + B_0 + B_1, 
\end{align*}
where 
\begin{align*}
&f_n(v):=\mu_n(e^{-v}-1)\\&\qquad \ \ +\frac{\mu_n}{\sum_{k\neq i,j }^{m} \mu_k} \Big(\mu_i(e^{-v}-1)+\lambda(e^v-1) C_0\Big)\\
     &\qquad \ \ +\frac{\mu_n}{\sum_{k\neq i,j}^{m} \mu_k} \Big(\mu_j(e^{-v}-1)+\lambda(e^v-1)\Big),\ \ \ \quad \forall n,\\
 &B_1=\sum_{n\neq i,j}^{m} \frac{\mu_n}{\sum_{k\neq i,j }^{m} \mu_k} \Big(\mu_i(e^{-v}-1)+\\&\qquad \qquad \ \ \lambda C_0(e^v-1) e^{-|x|_1} \Big)(e^{vx_{\max}}-e^{vx_{n}})                         \\&\ \quad
 \leq\sum_{n\neq i,j}^{m} \frac{\mu_n}{\sum_{k\neq i,j }^{m} \mu_k} \Big(\mu_i(e^{-v}-1)(e^{vx_{\max}}-e^{vx_{n}})+\\& \qquad\qquad \ (\sum_{k=1}^m \mu_k-\lambda)(e^v-1)(e^{vx_{\max}-|x|_1}-e^{vx_{n}-|x|_1}) \Big) .  
\end{align*}
Note that $B_1$ is finite as long as 
$v\leq 1$.
Note that $f_n$ is continuous and $f_n(0)=0$, $f_n(\infty)=\infty$. The derivative of $f(v)$ at $v=0$ is calculated as
\begin{align*}
    \frac{df_n}{dv}\Big|_{v=0}=\mu_n\Big(\frac{\lambda(1+C_0)-\mu_i-\mu_j}{\sum_{k\neq i,j }^{m} \mu_k} -1\Big) <0.
\end{align*}
Then the fact that derivative of $f_n(v)$ is negative at $0$ implies that there exist $v_0>0$ as the second zero of $f_n(v)$ such that $f_n(v)<0, v\in(0,v_0)$. Hence, we can guarantee (\ref{condition}) with a proper selection of $v^*\in(0,\min \{v_0,  1\}) $. The corresponding $c=-\max_n f_n(v^*)$, $d=B_0+B_1$ by \cite[Theorem 7.1]{meyn1993stability}.$\quad \qedsymbol$

\end{appendix}

\section*{Acknowledgments}
The authors appreciate the inputs from Qian Xie, Yidan Wu, Yule Zhang, and other members of the Smart \& Connected Systems Lab at Shanghai Jiao Tong University.

\section*{References}{
}
\vspace{-1.3\baselineskip}  
\bibliographystyle{IEEEtran}
\bibliography{Bibliography}

\begin{IEEEbiography}[{\includegraphics[width=1in,height=1.25in,clip,keepaspectratio]{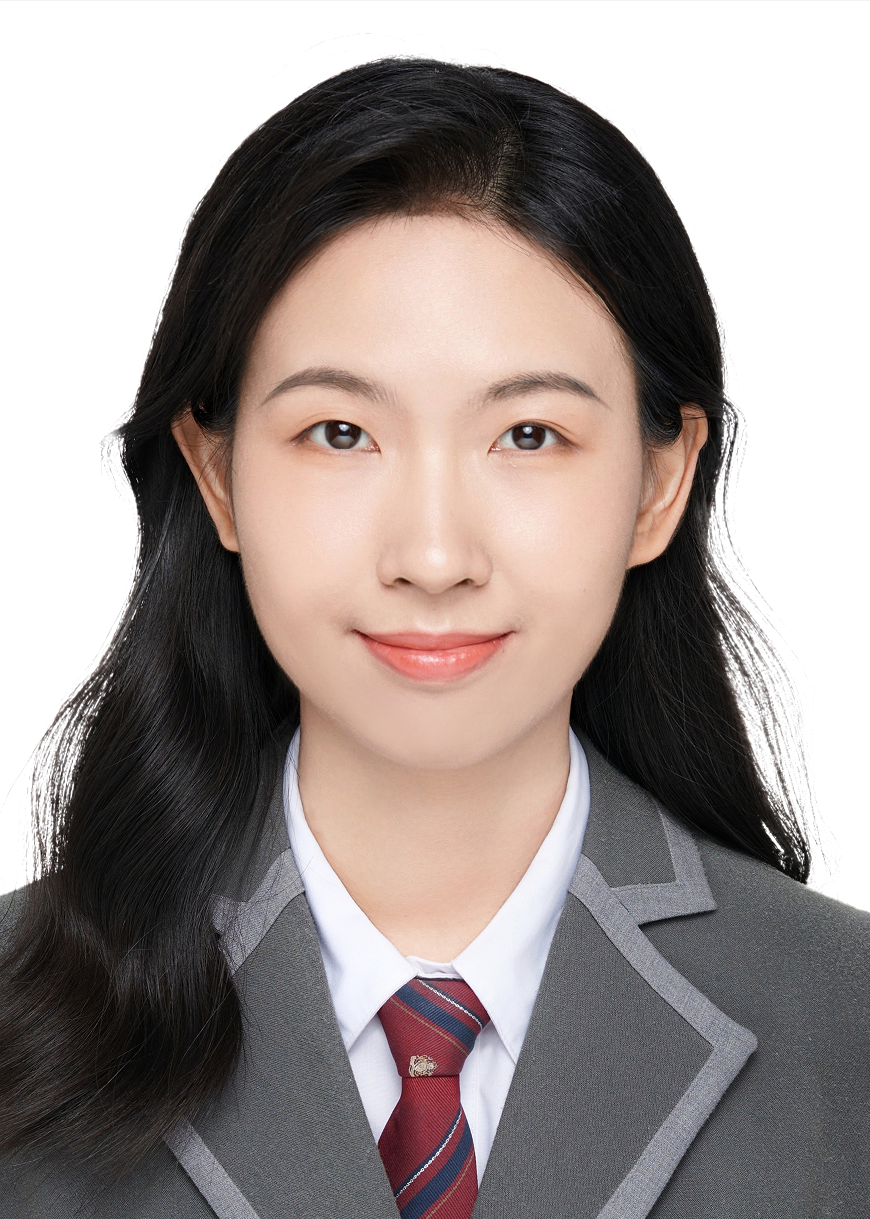}}]{Yuzhen Zhan}
is a master student (2023–present) at the UM Joint Institute, Shanghai Jiao Tong University, China. 
 She received her B.Eng. degree in Automation
 from Wuhan University, China in 2023.
 Her research focuses on game-theoretical model of adversarial dynamic and applying reinforcement learning methods to solve practical security challenges in cyber-physical systems. In particular, she is interested in theoretical guarantees of algorithms.
\end{IEEEbiography}


\begin{IEEEbiography}[{\includegraphics[width=1in,height=1.25in,clip,keepaspectratio]{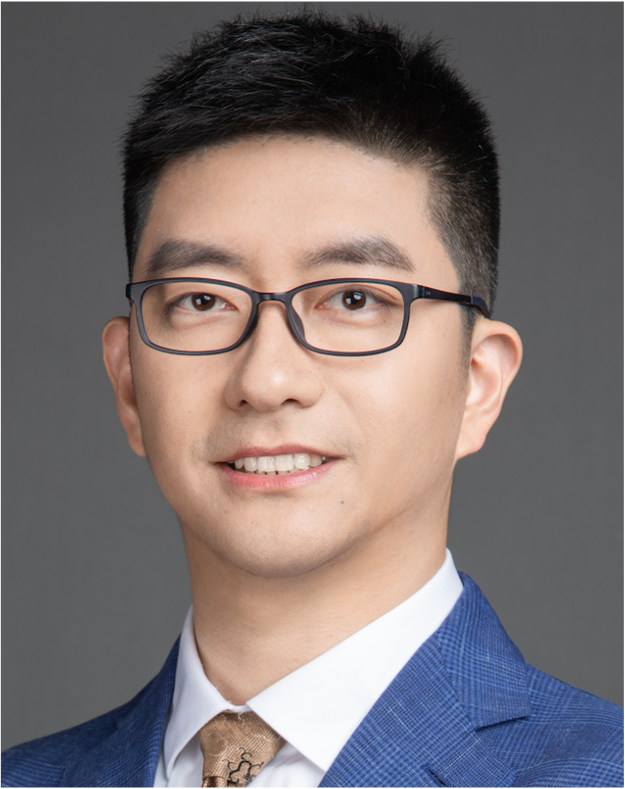}}]{Li Jin} is John Wu \& Jane Sun Associate Professor (2022--present) and was Assistant Professor (2021-2022) of Electrical and Computer Engineering at the UM Joint Institute and the Department of Automation, Shanghai Jiao Tong University (SJTU), China. He was Assistant Professor (2018--2020) at the Tandon School of Engineering, New York University, USA. He received his B.Eng. from SJTU in 2011, M.S. from Purdue University, USA in 2012, and Ph.D. from the Massachusetts Institute of Technology, USA in 2018. He was also a Visiting Scholar at the University of Erlangen-Nuremberg, Germany in 2016. 
He is the recipient of multiple research grants/awards from the US National Science Foundation and the National Natural Science Foundation of China on topics including connected and autonomous vehicles, network system control, and cyber-physical security.
\end{IEEEbiography} 
\end{document}